\makeatletter
\declare@file@substitution{revtex4-1.cls}{revtex4-2.cls}
\makeatother

\documentclass[twocolumn]{aastex631}

\usepackage{graphicx}	
\usepackage{amsmath}	
\usepackage[T1]{fontenc}
\usepackage{newtxtext,newtxmath}

\received{}

\submitjournal{ApJ}

\begin{document}

\title{Geometry of the comptonization region of MAXI~J1348--630 through type-C quasi-periodic oscillations with \textit{NICER}}

\author{Kevin Alabarta}
\affiliation{Center for Astrophysics and Space Science (CASS), New York University Abu Dhabi,\\ PO Box 129188, Abu Dhabi, UAE}

\author{Mariano M\'endez}
\affiliation{Kapteyn Astronomical Institute, University of Groningen,\\ PO Box 800, NL-9700 AV Groningen, the Netherlands}

\author{Federico Garc\'ia}
\affiliation{Instituto Argentino de Radioastronom\'ia (CCT La Plata, CONICET; CICPBA; UNLP), C.C.5, (1894) \\ Villa Elisa, Buenos Aires, Argentina}

\author{Diego Altamirano}
\affiliation{School of Physics and Astronomy, University of Southampton, Southampton, SO17 1BJ, UK}

\author{Yuexin Zhang}
\affiliation{Center for Astrophysics | Harvard \& Smithsonian, 60 Garden Street, Cambridge, MA 02138, USA}
\affiliation{Kapteyn Astronomical Institute, University of Groningen,\\ PO Box 800, NL-9700 AV Groningen, the Netherlands}

\author{Liang Zhang}
\affiliation{Key Laboratory of Particle Astrophysics, Institute of High Energy Physics, Chinese Academy of Sciences, Beijing 100049, China}

\author{David~M.~Russell}
\affiliation{Center for Astrophysics and Space Science (CASS), New York University Abu Dhabi,\\ PO Box 129188, Abu Dhabi, UAE}

\author{Ole K\"{o}nig}
\affiliation{Center for Astrophysics \textbar Harvard \& Smithsonian, 60 Garden Street, Cambridge, MA 02138, USA}
\affiliation{Dr. Karl Remeis-Observatory and Erlangen Centre for Astroparticle Physics, Friedrich-Alexander-Universität Erlangen-Nürnberg,\\ Sternwartstr. 7, 96049 Bamberg, Germany}



\begin{abstract}

We use the rms and lag spectra of the type-C quasi-periodic oscillation (QPO) to study the properties of the Comptonisation region (aka corona) during the low/hard and hard-intermediate states of the main outburst and reflare of MAXI~J1348$-$630. We simultaneously fit the time-averaged energy spectrum of the source and the fractional rms and phase-lag spectra of the QPO with the time-dependent Comptonization model \textsc{vkompth}. The data can be explained by two physically connected coronae interacting with the accretion disc via a feedback loop of X-ray photons. The best-fitting model consists of a corona of $\sim$10$^3$ km located at the inner edge of the disc and a second corona of $\sim$10$^4$ km horizontally extended and covering the inner parts of the accretion disc. The properties of both coronae during the reflare are similar to those during the \textit{low/hard} state of the main outburst, reinforcing the idea that both the outburst and the reflare are driven by the same physical mechanisms. We combine our results for the type-C QPO with those from previous work focused on the study of type-A and type-B QPOs with the same model to study the evolution of the geometry of the corona through the whole outburst, including the reflare of MAXI~J1348$-$630. Finally, we show that the sudden increase in the phase-lag frequency spectrum and the sharp drop in the coherence function previously observed in MAXI~J1348$-$630 are due to the type-C QPO during the decay of the outburst and can be explained in terms of the geometry of the coronae.

\end{abstract}

\keywords{accretion, accretion discs -- black hole physics -- stars: black holes -- X-rays: binaries -- X-rays: individual (MAXI J1348-630)}


\section{Introduction} \label{sec:intro}

During an outburst, black hole low-mass X-ray binaries (BH~LMXBs) show evolving spectral and timing properties \citep[e.g., ][]{VanDerKlis89,Mendez97,VanDerKlis00,Homan05b,Remillard06,Belloni10b,Belloni11,Plant14,Motta16}. Taking into account these properties, two main spectral states are defined \citep[see e.g., ][]{Tanaka89, vanderKlis94}: the low/hard state (LHS) and the high/soft state (HSS). During the LHS, the X-ray emission of the system is dominated by a Comptonized component (hereafter corona) described by a hard power-law in the X-ray energy spectrum. In the HSS, the X-ray emission is dominated by the accretion disc and described by a multi-colour disc blackbody. Between the LHS and the HSS, it is possible to distinguish two intermediate states \citep[][]{Homan05b}: the \textit{hard-intermediate state} (HIMS) and the \textit{soft-intermediate state} (SIMS), showing properties in between the LHS and HSS.

In some BH~LMXBs, at the end of the decay of an outburst, the X-ray luminosity increases again, reaching only up to 1$-$10\% of the luminosity at the peak of the outburst. These events are known as \textit{``reflares''} or \textit{``rebrightenings''} \citep[e.g., ][]{Callanan95, Chen97, Altamirano11a, Jonker12, Patruno16, Cuneo20, Zhang19, Zhang20, Zhang20a, Saikia23}. The similarities between the properties of BH~LMXBs during outbursts and reflares suggest that both types of events are driven by the same physical processes \citep[e.g., ][]{Patruno16, Cuneo20, Alabarta22}, although this is still not clear.

BH~LMXBs show strong variability that evolves along the different spectral states. During the LHS, the power-density spectrum (PDS) of the system is characterised by strong broadband noise with a fractional rms amplitude of 30\%--50\% \citep[e.g.][]{Mendez97,Belloni05,Remillard06, Munoz-Darias11, Motta16}. In contrast, in the HSS, the broadband fractional rms of BH~LMXBs is generally less than 5\% \citep{Mendez97}.

Low-frequency quasi-periodic oscillations (LF~QPO) can also be detected in the PDS of a BH~LMXB during outburst, with frequencies ranging from a few mHz to 30 Hz \citep[e.g., ][]{Psaltis99, Nowak00, Belloni02a, Casella04, Belloni05, Remillard06c, Belloni10}. Based on the combined properties of the QPO and the broadband noise components, these LF~QPOs are classified into type-A, -B and -C \citep[e.g., ][]{Wijnands99, Casella04, Casella05}. The type-C QPOs, which are the main point of this paper, consist of a strong and narrow peak with a centroid frequency ranging from 0.01 Hz to 30 Hz superposed to strong, 15--30\% fractional rms amplitude, broadband noise component that can be decomposed into individual Lorentzians \citep[e.g., ][]{Casella04, Belloni05}. Type-C QPOs are detected both in the LHS and the HIMS.


The radiative origin of type-C QPOs can be determined by the study of their energy-dependent timing properties, such as the fractional rms amplitude \citep[e.g., ][]{Tomsick01c, Casella04, Casella05, Rodriguez04, Rodriguez04a, Sobolewska06, Axelsson16, Zhang17b, Zhang20b, Karpouzas21, Ma21} and the lags between different energy bands \citep[e.g., ][]{Vaughan94, Vaughan97, Reig00, Cui00a, Casella04, Munoz-Darias10, Pahari13, Zhang17b, Jithesh19, Zhang20b, Mendez24}.

The fractional rms amplitude of type-C QPOs increases with energy up to 10--20 keV and remains more or less constant above that energy and up to $\sim$200 keV \citep[e.g., ][]{Casella04, Zhang17b, Zhang20b, Ma21}. The high fractional rms values of type-C QPOs at energies above $\sim$20--30 keV indicates that the corona dominates the QPO emission, supporting models that identify type-C QPOs as oscillations in the physical properties of the corona \citep[e.g., ][]{Lee01,Kumar14, Karpouzas20, Bellavita22}. 

Lags represent the delay between photons of two different energy bands. The lags are obtained from the Fourier cross-spectrum (CS) between the light curves in the subject and reference energy bands \citep[e.g, ][]{vanderklis96, Vaughan97, Nowak99}. If the high-energy photons are delayed with respect to those from low-energy bands, the lags are defined as positive and are called \textit{hard lags}. On the contrary, if the low-energy photons are delayed with respect to those from high-energy bands, the lags are defined as negative and are called \textit{soft lags}. Both soft and hard lags have been observed for type-C QPOs of BH~LMXBs \citep[e.g., ][]{Reig00, Cui00a, Casella04, Munoz-Darias10, Pahari13, Zhang17b, Jithesh19, Zhang20b}.

The physical origin of these lags is still poorly understood, although several mechanisms have been proposed to explain them. Hard lags, for instance, could be produced by Comptonization of soft photons from the accretion disc in the corona \citep[e.g., ][]{Payne80, Miyamoto88, Nobili00, Lee01, Kumar14, Karpouzas20}, or propagating fluctuations of the mass accretion rate from the outer parts of the accretion disc to the inner parts and the corona \citep[e.g., ][]{Arevalo06,Ingram13}. Soft lags, on the other hand, can be explained by the reverberation of photons from the corona impinging back in the accretion disc \citep[e.g., ][]{Uttley14, DeMarco15, Ingram20}. Reverberation, however, cannot explain the lags of QPOs \citep{DeMarco15}. Alternatively, soft lags can be produced by Comptonization if one takes into account feedback between the corona and the accretion disc \citep[][]{Lee01}, in which a fraction of the up-scattered photons in the corona impinges back onto the accretion disc producing a time delay between the hard photons from the corona and those from the accretion disc \citep{Karpouzas20, Bellavita22}.

In addition to the lags, the CS also allows us to study the coherence function, which represents a measure of the linear correlation between the photons coming from the hard and the soft bands \citep{Vaughan97}. Generally, in BH~LMXBs, the coherence is consistent with $\sim$1 in a broad range of frequencies, indicating that hard photons are strongly related to soft photons \citep[e.g, ][]{Vaughan97, Cui97, Nowak99, Ji03}. However, some drops in the coherence function at certain frequency ranges have been observed \citep[e.g., ][]{Cui97, Nowak99, Ji03, Mendez24, Konig24}. \cite{Konig24} found a drop in the coherence function in \textit{NICER} observations of Cyg~X$-$1 at frequencies in between the two broad Lorentzian components describing the PDS of the system without finding a significant feature at the frequency of the drop in the PDS. These authors speculated that this drop was caused by a beat between the two broad Lorentzians. Alternatively, \cite{Mendez24} observed a similar feature in the coherence function of MAXI~J1820$+$070. Fitting the PDS and CS with the same multi-Lorentzian function, they found a Lorentzian only significant in the Imaginary part of the CS (called ``Imaginary QPO'' by \cite{Mendez24}) at the same frequency as the drop in coherence. This suggested that both phenomena were produced by the same independent physical mechanism.

The X-ray timing properties of LMXBs also allow the study of the geometry of the corona and its evolution. As a result, recently, it has been shown that the size of the corona changes throughout the outburst \citep[e.g., ][]{Kara19, Wang21, Garcia22, Mendez22, Zhang23}. Several models have been proposed to explain the evolution of the geometry of the corona. An outflowing corona model was proposed to explain the correlation of the photon index characterising the corona with the broadband time lags and radio flux \citep[e.g., ][]{Kylafis08, Reig18, Kylafis18, Kylafis23}. Alternatively, the corona has also been proposed to be the base of a jet emitted from the centre of the system \citep{Markoff05}. The propagating mass accretion rate fluctuations assume a corona that lies over the inner parts of the accretion disc and is able to explain the broadband and QPO variability, as well as the energy spectra of LMXBs \citep[][]{Arevalo06, Ingram13, Zdziarski21, Kawamura22}. The family of \textsc{reltrans} models assume a lamppost geometry for the corona and measure its height with respect to the accretion disc by modelling the broadband lags due to reverberation \citep{Ingram19, Lucchini23}. The JED-SAD model \citep{Ferreira97, Ferreira06_jed1, Petrucci08, Marcel18_jed2, Marcel18_jed3, Marcel19_jed4, Marcel20_jed5, Marcel22_jed6} assumes that the hard component of the X-ray emission can be described by a jet-emitting disc (JED) and the soft component by a standard accretion disc (SAD). This model explains the spectral and timing properties of LMXBs by changes in the geometry and the contribution of the JED and SAD components to the total X-ray emission. 

Because they assume a monolithic time- and energy-dependent response function of the system \citep[e.g., due to the accretion disc, ][]{Reynolds99}, all these models consider that the time lags are the result of a single process that leads to a single broadband variability component, the so-called broadband noise component, in the PDS and the CS of these sources. \cite{Mendez24}, however, have recently shown that, as was already the case for the PDS \citep[][]{Nowak00, Belloni02a}, the CS can also be decomposed in a number of Lorentzian functions (in this case, each of them multiplied by the cosine or sine of a frequency-dependent phase lag). This favours models of variability that consist of a number of response functions that act over well-defined time scales. This type of models includes Lense-Thirring precession \citep{Ingram09} and \textsc{vkompth} \citep{Karpouzas20, Bellavita22}.

This study focuses on the recently developed model \textsc{vkompth} \citep{Karpouzas20, Bellavita22}, which has been able to reproduce the rms and lag spectra of QPOs and measure the characteristic size of the corona around LMXBs \citep[e.g., ][]{Karpouzas20, Karpouzas21, Garcia21, Garcia22, Mendez22, Zhang22, Zhang23, Rawat23, Rout23, Ma23}. The \textsc{vkompth} model considers that the coupled oscillation between the corona and the accretion disc, at the frequency of the QPO, produces the radiative properties of QPOs. This model allows us to link the behaviour of these properties with the coronal geometry of the system.


MAXI~J1348--630 is an X-ray binary discovered with MAXI in January 2019 \citep[][]{Yatabe19, Tominaga20}. The source was also detected with \textit{Swift} \citep[][]{D'Elia19a, D'Elia19b}, \textit{NICER} \citep[][]{Sanna19}, \textit{INTEGRAL} \citep[][]{Cangemini19a}, \textit{ATCA} \citep[][]{Russell19}, \textit{iTelescope.Net} \citep[][]{Denisenko19} and LCO \citep[][]{Russell19_1348}. Based on its spectral and timing properties studied with \textit{NICER} data, \citet{Sanna19} suggested that the compact object in this system is a BH. Later on, a more detailed study of the evolution of these properties during the whole outburst using \textit{NICER} allowed to reinforce the identification of MAXI~J1348--630 as a BH candidate \citep[][]{Zhang20}.

In this paper, we present the evolution of the properties of the corona of MAXI~J1348$-$630 using the version of the Comptonization model \textsc{vkompth} with two coronas, \textsc{vkdualdk} \citep{Garcia21, Bellavita22} on the type-C QPO detected in this system. In Section 2, we describe the observations and data analysis. In section 3, we show the best-fit parameters of the observations included in this study and present the parameters describing the corona in the different phases of the outburst. Finally, in section 4, we compare the coronal properties of MAXI~J1348$-$630 in the main outburst with those in the reflare and with other LMXBs and from them, we infer the potential geometry of the two coronae. Finally, we also show the evolution of the geometry based on the studies of the three types of QPOs observed in MAXI~J1348$-$630.

\section{Observations \& data analysis}

\cite{Alabarta22} published a complete timing analysis of the \textit{NICER} observations in which a type-C QPO was observed. They analysed 37 observations with type-C QPOs and gave the ObsIDs and the properties of the QPO in their Table 1. In this section, we briefly describe the data processing and refer the reader to \cite{Alabarta22} for the details. 

\subsection{Light curve and HID}

\begin{figure*}
    \centering
    \includegraphics[width=\textwidth]{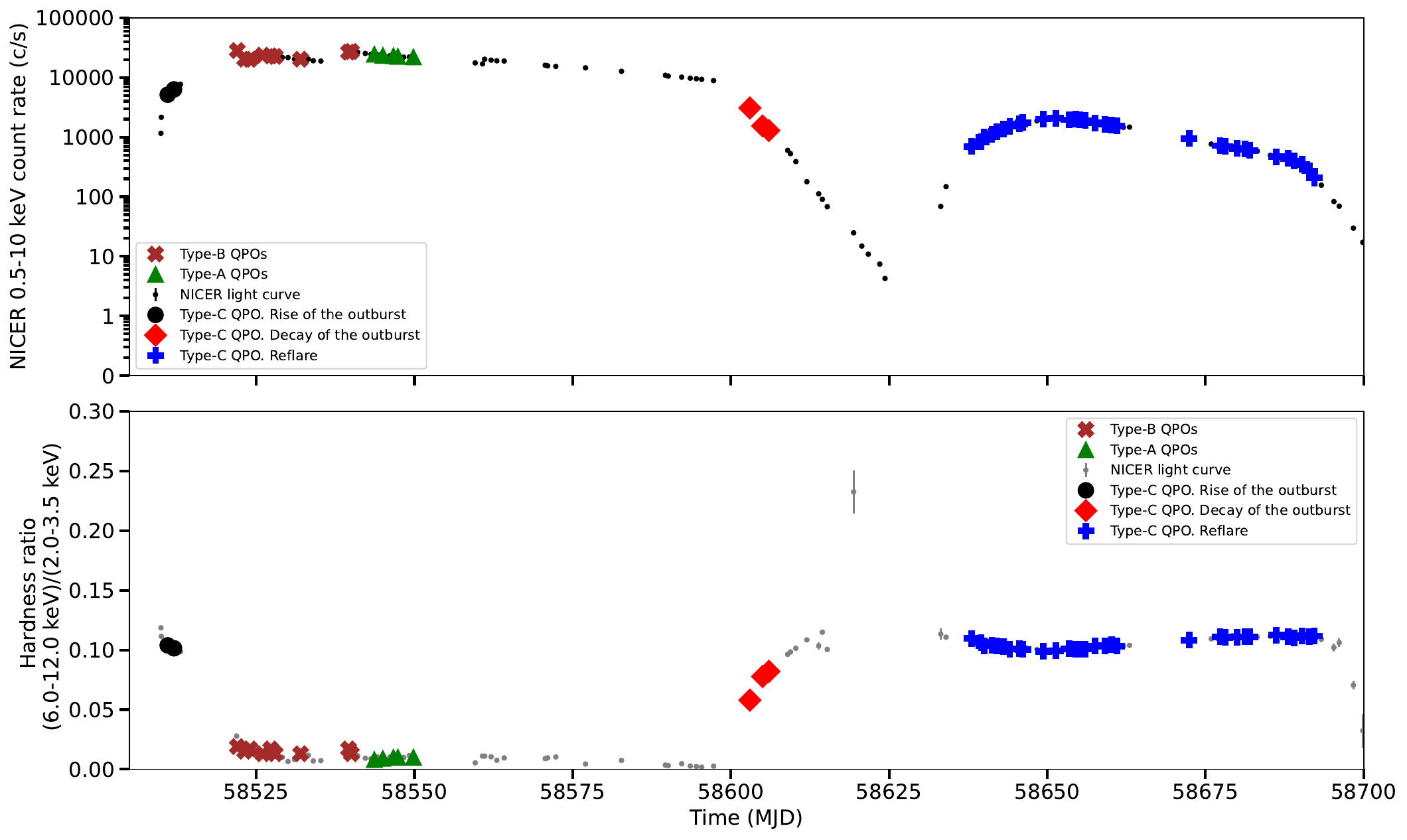}
    \caption{\textit{NICER} light curve (top panel) and hardness ratio vs. time (bottom panel) of MAXI~J1348$-$630. Black, red and blue symbols represent the observations corresponding, respectively, to the rise, decay and first reflare of the outburst in which type-C QPOs were observed. Brown and green symbols represent, respectively, the observations in which type-B \citep{Belloni20} and type-A QPOs \citep{Zhang23_1348} were detected. Grey symbols represent the rest of the \textit{NICER} light curve (top panel) and the rest of the temporal evolution of the hardness ratio (bottom panel).}
    \label{fig:lc_hardness}
\end{figure*}

\begin{figure}
    \centering
    \includegraphics[width=0.9\columnwidth]{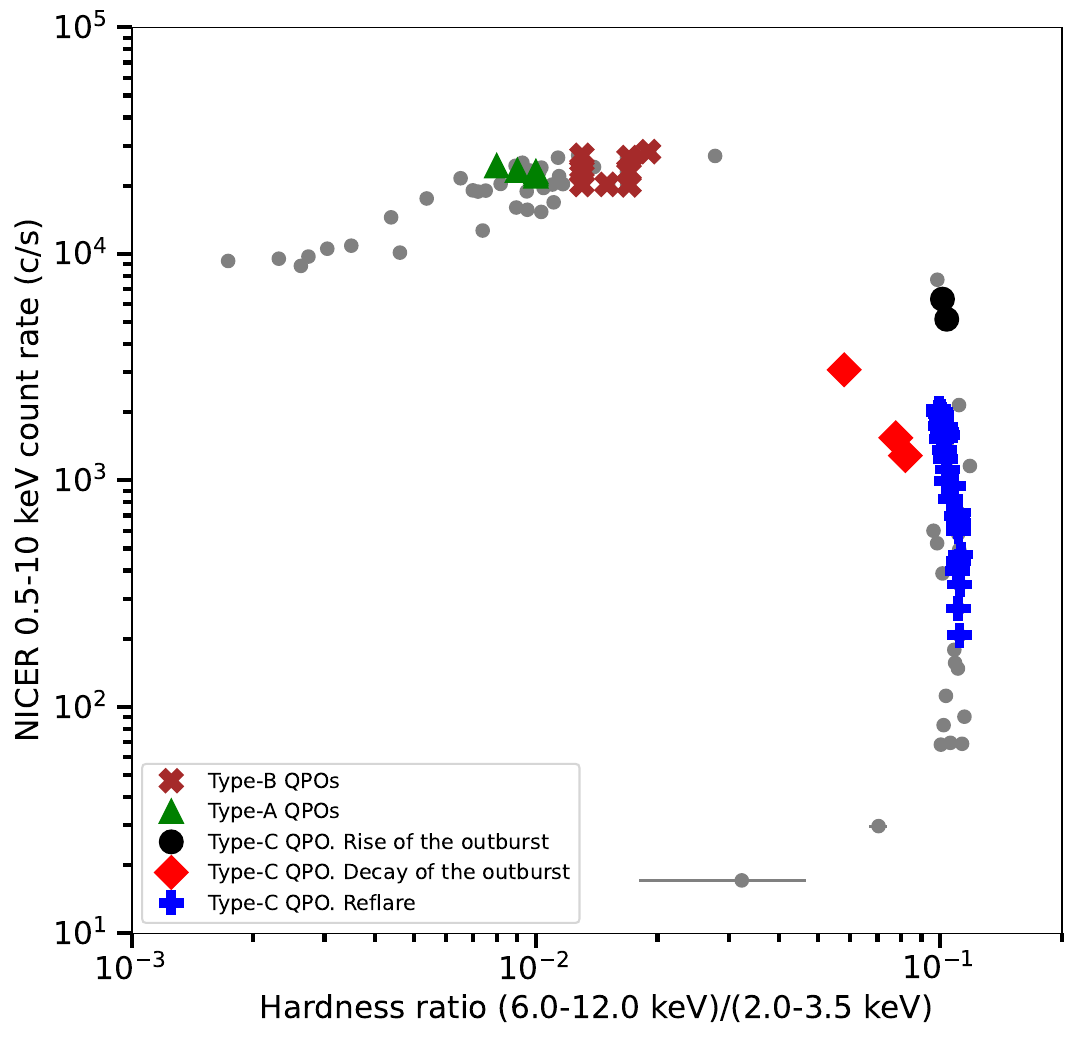}
    \caption{Hardness-intensity diagram (HID) of MAXI~J1348$-$630. Colours and symbols are the same as in Fig. \ref{fig:lc_hardness}.}
    \label{fig:hid_qpos}
\end{figure}

To recreate the light curve and HID presented in \citet{Alabarta22}, we extracted a background-subtracted energy spectrum for each ObsID using the \textsc{nibackgen3c50} background model \citep{Remillard22}. We obtained the count rate of the source in the 0.5--12 keV, 2--3.5 keV and 6--12 keV energy bands for each ObsID, defined the intensity as the background-subtracted count rate in the 0.5--12 keV energy range and the hardness ratio as the ratio between the background-subtracted count rate of the 6--12 keV and 2--3.5 keV energy bands (one point per ObsID). Figure \ref{fig:lc_hardness} shows the \textit{NICER} light curve and hardness ratio vs time, and Figure \ref{fig:hid_qpos} shows the HID of MAXI~J1348$-$630. In both figures, we marked the observations with different symbols, showing type-A, -B, and -C QPOs.

\subsection{Time-averaged energy spectra}

We fitted the energy spectra of MAXI~J1348--630 in the energy band $0.7-10$ keV using XSPEC \citep[V. 12.10.1;][]{Arnaud96}. We rebinned the spectra by a factor of 3 to correct for energy oversampling and then to have at least 25 counts per bin. In addition, we added a systematic error of 1\% in the energy range $0.7-10$ keV. We found strong instrumental residuals below 3 keV. These residuals are typical for X-ray missions with Si-based detectors \citep[e.g.][]{Ludlam18a,Miller18}. We, therefore, fitted those residuals with three \textsc{edge} and one \textsc{gabs} components, and ignored the $1.7-2.3$ keV energy range. We fitted the energy spectra with an absorbed \citep[\textsc{tbfeo} in XSPEC, ][]{Wilms00} combination of a thermal Comptonisation model \citep{Zdziarski96, Zycki99} and a multi-colour disc blackbody, \textsc{tbfeo(nthcomp+diskbb)}. In order to obtain the fluxes of the different components, we added two \textsc{cflux} components to the models. The solar abundances were set according to \citet{Wilms00}, and the hydrogen column density ($N_{\rm H}$) of the \textsc{tbfeo} was left free. The cross-section was set according to \citet{Verner96}. The 1$\sigma$ errors of the parameters were calculated from a Monte Carlo Markov Chain of length 240000 with a 2400-step burn-in phase.

First, we fitted all the energy spectra separately, linking the $kT_{bb}$ parameter of \textsc{nthcomp} and $kT_{in}$ of \textsc{diskbb}. We found that the electron temperature, $kT_{e}$, in \textsc{nthcomp} was always above the maximum energy of the instrument. Therefore, we fixed $kT_{e}$ at 50 keV. Besides, we noted that the values of the parameters of \textsc{tbfeo} in all the fitted energy spectra were consistent within errors. For this reason, we decided to fix $N_{H}$ to 0.86$\times 10^{22}$ cm$^{-2}$ as in \citet{Carotenuto21}, link the parameters $O$ and $Fe$ in all the spectra and to repeat the fitting.

The fit with this model shows significant residuals at 6$-$7 keV. These residuals are thought to be due to the reflection component \citep[e.g., ][]{Garcia14, Fabian89}. Therefore, we added to the model the relativistic reflection component, \textsc{relxillCp} \citep[version 1.4.3; ][]{Dauser14}). It is necessary to point out that the seed photons in \textsc{relxillCp} correspond to the Comptonized \textsc{nthComp} spectrum of a disc-blackbody with a fixed temperature of $kT_{in}$ = 0.01 keV. This spectrum is very different from that of LMXBs, in which the typical disc-blackbody temperatures range between 0.5 keV and 2 keV. This introduces a bias in the spectral modelling, producing a soft excess at $\sim$2 keV. In order to correct for this soft excess, we normalised this component by the ratio between an \textsc{nthComp} model evaluated at $kT_b$ = $kT_{in}$ and the \textsc{nthComp} at 0.05 keV, using the so-called \textsc{nthratio} model\footnote{\url{https://github.com/garciafederico/nthratio}}. We note that \textsc{nthratio} takes no extra degrees of freedom because all of its parameters are linked to the corresponding parameters of \textsc{nthComp}. We linked $kT_e$ and $\Gamma$ in \textsc{relxillCp} with the corresponding parameters $kT_e$ and $\Gamma$ in \textsc{nthComp}. We then fixed the BH spin, $a_{\ast}$, at 0.998. In all the observations, we also linked the inclination, $i$, and iron abundance, $A_{Fe}$. In order to get only the reflected emission of \textsc{relxillCp}, we fixed the reflection fraction to –1. We set the disc inner radius, $R_{in}$, at the ISCO, linked the emissivity indices of the relxillCp, $\alpha_1$, and $\alpha_2$, to be the same, and we let the ionisation parameter, $\xi$, and the normalisation of \textsc{relxillCp} free.

\subsection{PDS, rms and lag spectra, and coherence function}

The representative PDS of the observations of MAXI~J1348--630 were presented in Figure 3 of \citet{Alabarta22} and the properties of the type-C QPO were shown in Table 1 of \citet{Alabarta22}. All the PDS were fitted with a multi-Lorentzian function, using the characteristic frequency of the Lorentzians defined in \citet{Belloni02a}, $\nu_{max}=\sqrt{\nu_{0} + (FWHM/2)^2} = \nu_{0} \sqrt{1+1/4Q^{2}}$, where $FWHM$ is the full width at half maximum, $\nu_{0}$ the centroid frequency of the Lorentzian and the quality factor $Q$ is defined as $Q = \nu_{0}/FWHM$. In \cite{Alabarta22}, we followed these criteria to identify the QPO: if there was only one narrow component, we identified it as the QPO. If there were two or more narrow Lorentzians, we assumed that the fundamental QPO was the strongest and narrowest Lorentzian and if the other narrow Lorentzians were at frequencies $\sim$1/2 or $\sim$2 times the frequency of the QPO, these Lorentzians were identified as the subharmonic or the harmonic, respectively.    

To obtain the rms spectrum, \cite{Alabarta22} carried out the procedure described above in the following energy bands: $0.5-1.0$ keV, $1.0-1.5$ keV, $1.5-2.0$ keV, $2.0-3.0$ keV, $3.0-4.0$ keV, $4.0-6.0$ keV, $6.0-8.0$ keV and $8.0-12.0$ keV. 

In order to compute the phase-lag spectrum of the Lorentzian associated with the type-C QPO, \cite{Alabarta22} produced the CS using the selected subject bands listed above and 0.5$-$12 keV as the reference band. They fit simultaneously the Real and Imaginary parts of each cross-spectrum in \textsc{XSPEC} using the multi-Lorentzian model used to fit the PDS, with the centroid frequencies and FWHM of each Lorentzian in the model fixed, letting their normalisations vary. Next, the phase-lag of the fundamental QPO was calculated by taking the arctan of the ratio of the normalization of the Lorentzians associated with the Imaginary and Real parts of the QPO, respectively, $\phi = \mathrm{atan(Im(QPO)/Re(QPO))}$.

This method has also been used in \citet{Peirano23,Zhang23,Ma23} and fully developed in \cite{Mendez24}. In particular, the analysis in \cite{Alabarta22} corresponds to the case of the constant phase-lag model in \cite{Mendez24}. To compute the broadband phase lags and coherence function vs. frequency in Section 3.3, we selected 0.5$-$2.0 keV and 2.0$-$10.0 keV as the reference and subject energy bands, respectively.

\subsection{Joint fitting of the time-averaged, rms and phase-lag spectra}

We fit the time-average spectra of the source, and the fractional rms and the phase-lag spectra of the type-C QPO simultaneously. For the energy spectra of the source, we used the model described in Section 2.2. For the fractional rms and phase-lag spectra of the QPO, we used the time-dependent Comptonization model \textsc{vkompth} presented in \cite{Karpouzas20} and \cite{Bellavita22}. This model assumes that the QPO is a small oscillation around the solution of the stationary Kompaneets equation \citep{Kompaneets57}, which describes the energy spectrum of the system. This model allows us to link the behaviour of the energy-dependent phase-lags and fractional rms amplitude with the physical properties of the system.      

Initially, we considered the version of \textsc{vkompthdk} for one corona. This model considers that the seed photon source is a geometrically thin, optically thick accretion disc \citep{Shakura73} with a temperature $kT_s$, surrounded by a corona of hot electrons with temperature $kT_e$, where the seed photons are inverse-Compton scattered. This corona is assumed to be spherically symmetric with a size $L$ and constant optical depth $\tau$. The model also includes feedback from the up-scattered photons in the corona to the seed photon source. This feedback is described by the parameter called ``feedback-fraction'', $\eta$, whose values range between 0 and 1, and is given by $$\eta = \dfrac{F_{\rm disc}-F_{\rm disc,o}}{F_{\rm disc}}$$
where $F_{\rm disc}$ is the total flux of the disc and $F_{\rm disc,o}$ represents the flux of the disc before the corona illuminates it. In summary, $\eta$ represents the fraction of the disc flux produced by the feedback process. The parameter $\eta$, in turn, is related to $\eta_{int}$ 
\citep[intrinsic feedback fraction; see ][]{Karpouzas20}, where $\eta_{int}$ is defined as the fraction of the coronal flux that illuminates the disc. In addition to these parameters, \textsc{vkompthdk} also considers an unspecified external heating source of the corona, $\delta \dot{H}_{ext}$, and an additive parameter, $reflag$, that gives the phase lag in the 2--3 keV band. We also added a \textit{dilution} correction factor (defined as \textsc{nthcomp/(nthcomp+diskbb+relxillCp)}) that takes into account the effect of the non-variable emission on the fractional rms amplitude. By doing this, we are assuming that all the variability of the system originates from the Comptonized component. 

During the fit, $kT_s$ $kT_{e}$ and $\Gamma$ of \textsc{vkompthdk} are linked to the corresponding parameters of \textsc{diskbb} ($kT_s$), and \textsc{nthcomp} ($kT_e$ and $\Gamma$). We found that fitting one corona cannot reproduce the rms and phase-lag spectra of MAXI~J1348$-$630. Because of this, we also fit the spectra with two Comptonization regions \citep[e.g., ][]{Garcia21,Bellavita22} using the model \textsc{vkdualdk} \citep{Bellavita22}. The two coronae (hereafter called corona 1 and corona 2) are assumed to be small and large, respectively, and described by two sets of parameters similar to those of the case of one corona: $kT_{s,1}$ and $kT_{s,2}$, $kT_{e,1}$ and $kT_{e,2}$, $L_{1}$ and $L_{2}$, $\eta_{1}$ and $\eta_{2}$, and $\delta \dot{H}_{ext,1}$ and$\delta \dot{H}_{ext,2}$. For simplicity, we assume that the photon index and $kT_e$ of both coronae are equal. On the other hand, $kT_{s,1}$ is linked to $kT_{in}$ on \textsc{diskbb} and $kT_{s,2}$ is set as a free parameter. The 1$\sigma$ errors of the parameters were calculated from a Markov Chain Monte Carlo of length 240000 with a 2400-step burn-in phase.

\begin{table}
\caption{Table of the \textit{NICER} observations analysed in this work. From observation 6, the ObsIDs within the interval in the second column were used to produce the energy, rms and lag spectra.}
\resizebox{\columnwidth}{!}{
\begin{tabular}{cccc}
\hline 
Obs number & ObsID & QPO freq. (Hz) & Phase of the outburst \\
\hline
\smallskip
1 & 1200530103 & 0.45$\pm$0.04 & Rise \\
\smallskip
2 & 1200530104 &0.53$\pm$0.01 & Rise \\ 
\smallskip
3 & 2200530127 & 2.93$\pm$0.04 & Decay \\ 
\smallskip
4 & 2200530128 & 1.67$\pm$0.06 & Decay \\ 
\smallskip
5 & 2200530129 & 1.56$\pm$0.06 & Decay \\ 
\smallskip
6 & 2200530145-2200530147 & 0.60$\pm$0.04 & Reflare \\ 
\smallskip
7 & 2200530148-2200530149 & 0.88$\pm$0.04 & Reflare \\ 
\smallskip
8 & 2200530151-2200530162 & 0.94$\pm$0.01 & Reflare \\ 
\smallskip
9 & 2200530165-2200530166 & 0.73$\pm$0.03 & Reflare \\ 
\smallskip
10 & 2200530171-2200530186 & 0.40$\pm$0.01 & Reflare \\ 
\hline
\end{tabular}}
\label{Tab:observations}
\end{table}

\section{Results}

\subsection{Fit to the energy spectra}

We first fitted simultaneously all the observations of MAXI~J1348$-$630 showing a type-C QPO (see Table. 1 in \cite{Alabarta22}). In order to increase the signal-to-noise ratio of the rms and phase-lags spectra of the QPO during the reflare, we combined all the observations with similar timing and spectral properties, resulting in a total of five energy spectra during the reflare of MAXI~J1348$-$630 (see Table \ref{Tab:observations}). We obtained a relatively good fit, with a $\chi^2/dof$ of 1.13, for 1884 degrees of freedom (for a total of 10 spectra).

The best-fitting parameters and their evolution are given in table \ref{Tab:spec_params}. As stated in Section 2, we fixed the $N_{H}$ at 0.86$\times 10^{22}$ $cm^{-2}$ and got for the oxygen and iron abundances of the interstellar medium along the line of sight $O$=0.8 $\pm$ 0.2 and $F_e$=0.34 $\pm$ 0.02. We obtained an inclination of $i$=49.445$^\circ$ $\pm$ 0.001$^\circ$, which is slightly higher than the estimations of previous studies \citep[e.g., ][]{Carotenuto22, Alabarta22}. 

The observations showing a type-C QPO during the rise of the outburst of MAXI~J1348$-$630 show $\Gamma$ $\sim$1.6 and an inner-disc temperature of $\sim$0.5 keV. During the HIMS, the photon index and the disc temperature decrease, respectively, from $\sim$2.5 to $\sim$2.0 and from $\sim$0.3 keV to $\sim$0.2 keV. Finally, during the reflare, the photon index remains constant at $\sim$1.7. In contrast, $kT_{in}$ increases from 0.24 keV at the beginning of the reflare to 0.36 keV at its peak and, after that, decreases until 0.25 keV at the last energy spectrum of the reflare.  

\subsection{Evolution of the parameters describing the corona}

\begin{figure*}
    \centering
    \includegraphics[width=\textwidth]{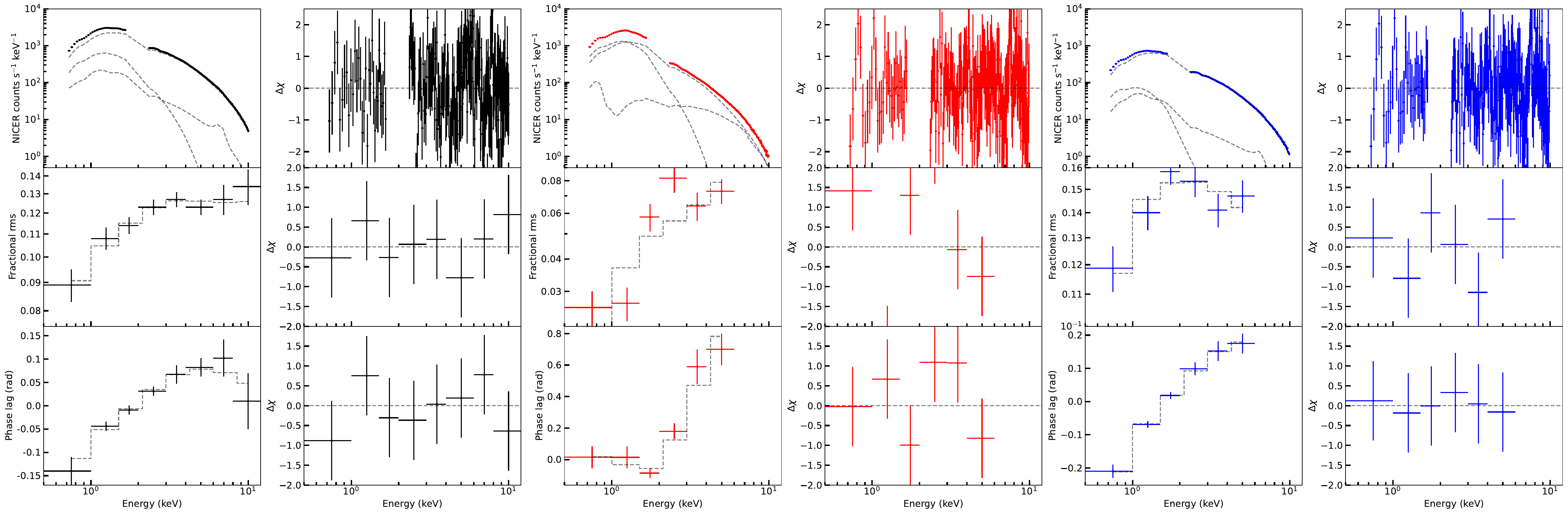}
    \caption{Three representative examples of the joint fitting of the time-averaged spectrum of the source (top), the fractional rms (middle) and lag spectra (bottom) of the type-C QPO during the rise (first column), decay (third column) and the reflare (fifth column) of MAXI~J1348$-$630. The second, fourth, and sixth columns show the residuals of the fits in the panel on the left side of each residual. Black, red, and blue symbols represent the data corresponding to the rise, decay and first reflare of the outburst, respectively.}
    \label{fig:spec_params}
\end{figure*}

\begin{figure*}
    \centering
    \includegraphics[width=\textwidth]{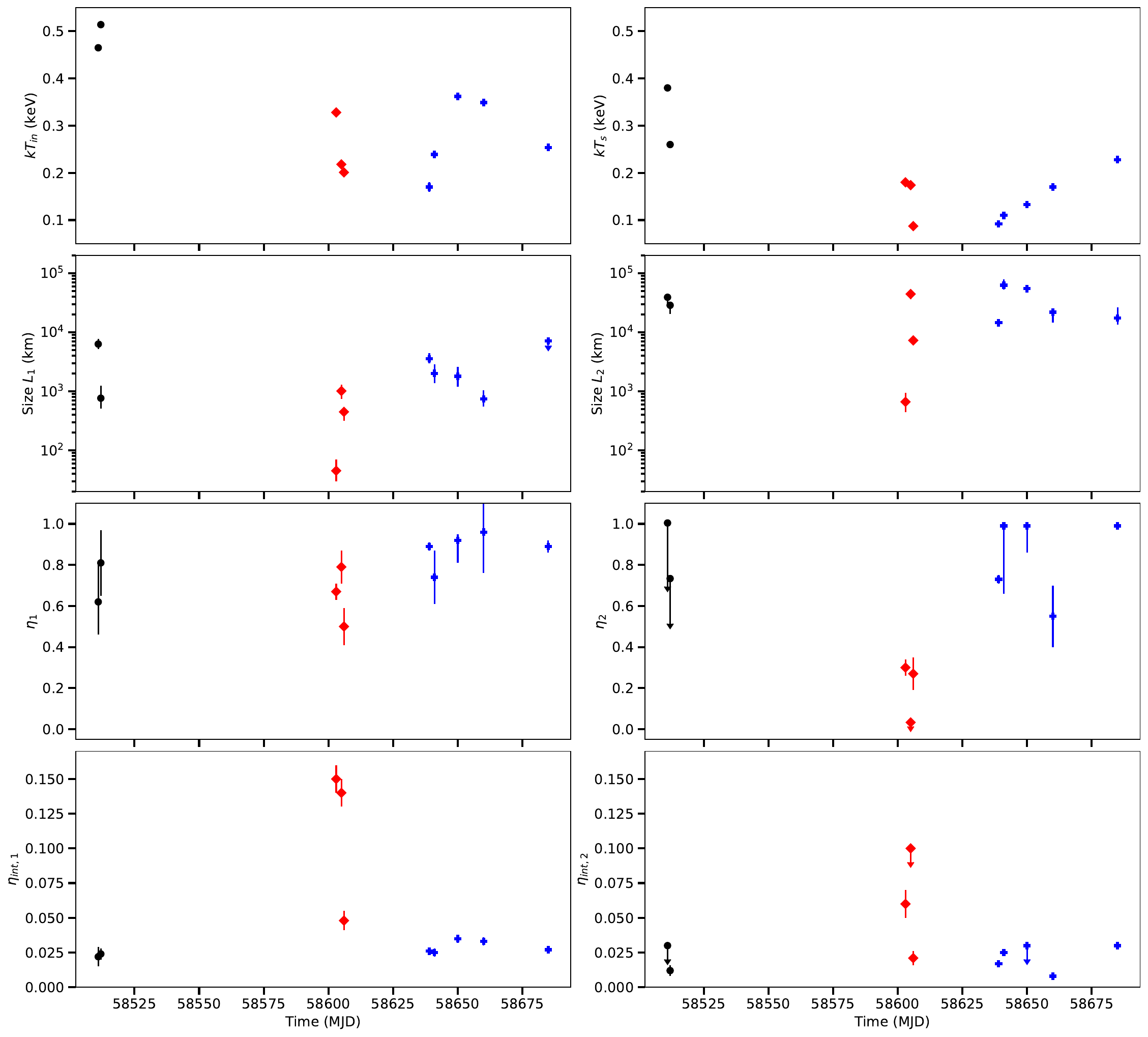}
    \caption{Evolution of the properties of the corona of MAXI~J1348$-$630. The different colours represent the different phases of the outburst as defined in the previous figures.}
    \label{fig:corona_params_time}
\end{figure*}

\begin{figure*}
    \centering
    \includegraphics[width=\textwidth]{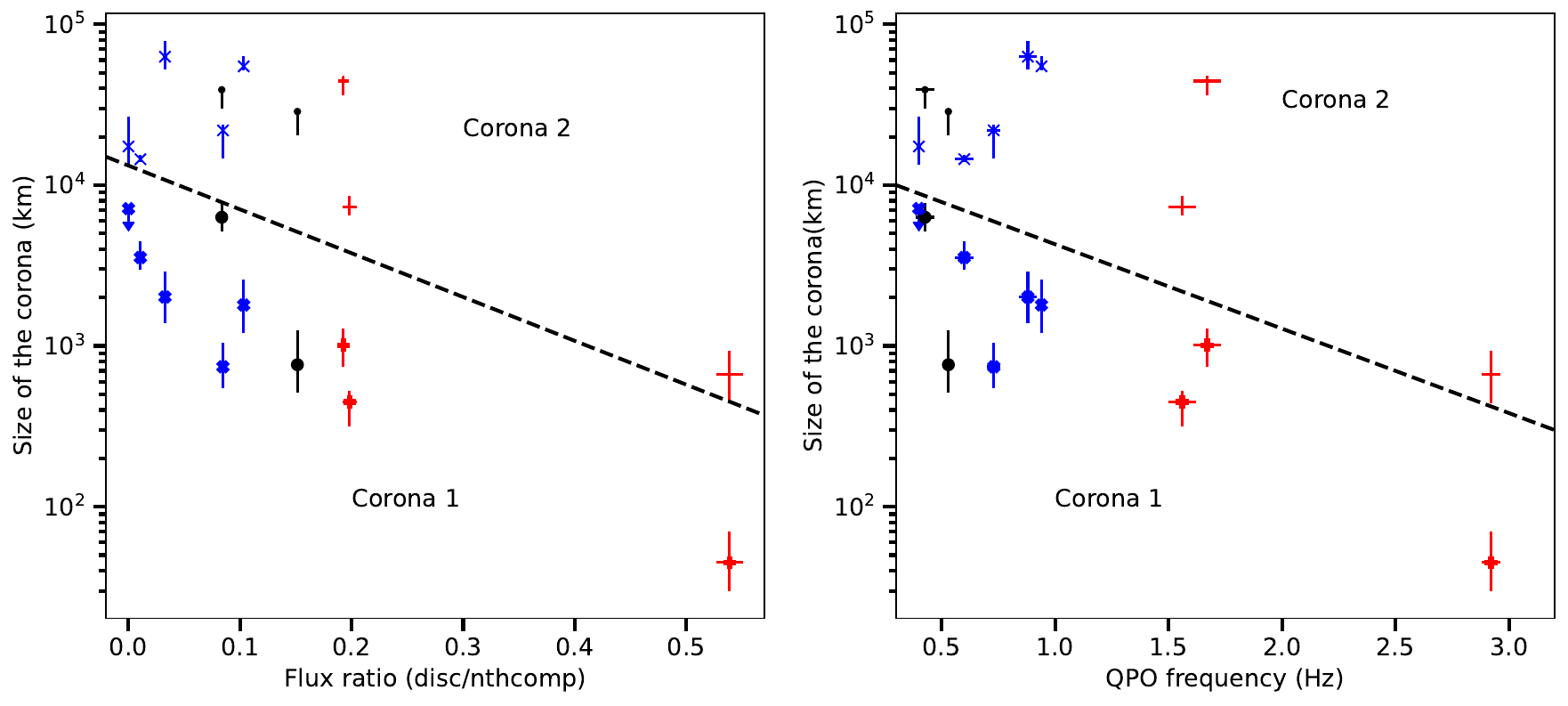}
    \caption{Evolution of the sizes of the small, $L_1$ and large corona, $L_2$, with the flux ratio (left panel) and the QPO frequency (right panel). The different colours represent the different phases of the outburst, as defined in the previous figures. The dashed line separates the data of both coronae.}
    \label{fig:corona_sizes_flux}
\end{figure*}

We then simultaneously fitted the energy spectrum of the source and the rms and phase-lag spectra of the QPO for each observation (or combination of observations, as explained in Section 3.1). 

Figure \ref{fig:spec_params} shows three representative joint fits of each phase of the outburst: rise and decay of the main outburst in, respectively, black and red, and the reflare in blue. The best-fitting parameters are given in Table \ref{Tab:spec_params}, and the time evolution of the parameters is shown in Figure \ref{fig:corona_params_time}. 

During the rise and the decay of the main outburst, $kT_{\rm s}$ decreases from 0.38 $\pm$ 0.03 keV to 0.087 $\pm$ 0.003 keV. During the reflare, $kT_{\rm s}$ initially increases from 0.092 $\pm$ 0.008 keV at the beginning to 0.17 $\pm$ 0.03 keV just after the peak. The last observation shows a slight decrease of $kT_{\rm s}$ to 0.13 $\pm$ 0.02 keV; however, the last two values are consistent within errors.

The size of the small corona, $L_{1}$, initially decreases from $\sim$6300 km ($\sim$425$R_{g}$, all sizes in $R_g$ units are for a BH of 10$M_{\odot}$) to $\sim$760 km ($\sim$51$R_{g}$) during the observations corresponding to the rise of the main outburst. During the decay, $L_1$ increases from $\sim$45 km ($\sim$3$R_{g}$) to $\sim$1000 km ($\sim$68$R_{g}$). Finally, during the reflare, $L_1$ decreases from $\sim$3550 km ($\sim$240$R_{g}$) in Obs 6 to $\sim$740 km ($\sim$50$R_{g}$) in Obs 9. In the last observation, the upper limit to the size of the small corona is $<$3000 km ($\sim$200$R_{g}$). The feedback fraction of this corona, $\eta_1$, ranges from 0.50 $\pm$ 0.09 to 0.96 $\pm$ 0.20. However, their relatively large errors make all the values consistent with each other. 

On the other hand, the size of the large corona, $L_2$, is $\sim$39000 km ($\sim$2600$R_{g}$) in Obs 1 and less than $\sim$55000 km ($\sim$3700$R_{g}$) in Obs 2, both during the rise of the main outburst. During the decay, $L_2$ increases from $\sim$660 km ($\sim$45$R_{g}$) in Obs 3 to $\sim$44000 km ($\sim$3000$R_{g}$) in Obs 4, while it decreases to $\sim$7300 km ($\sim$500$R_{g}$) in Obs 5. Finally, during the reflare, $L_2$ increases from $\sim$14500 km ($\sim$980$R_{g}$) to a value ranging between $\sim$52800 km ($\sim$3600$R_{g}$) and $\sim$63000 km ($\sim$4200$R_{g}$). Obs 9 shows an $L_2$ of $\sim$ 22000 km ($\sim$1500$R_{g}$), which is lower than that of the observations before and after. The feedback fraction of this corona, $\eta_2$, shows different values depending on the phase of the outburst. During the rise of the main outburst, we got only upper limits for $\eta_2$, $<$1.00 and $<$ 0.73, for Obs 1 and Obs 2, respectively. During the reflare, $\eta_2$ increases from $\sim$0.75 in Obs 6 to $\sim$1.00 in Obs 7,8 and 9, while in Obs 10, $\eta_2$ is 0.55. During the decay of the main outburst, the values for $\eta_2$ are lower than those during the rise of the main outburst and the reflare, ranging between $<$0.03 and $\sim$0.3. The values corresponding to the intrinsic feedback fraction are, for the small corona, $\eta_{int,1}$$\sim$0.03 in the rise and the reflare of the outburst and $\eta_{int,1}$$\sim$0.15 in the decay of the outburst. The parameter $\eta_{int,2}$, on the other hand, is $\sim$0.02 during the rise and the reflare and $\sim$0.07 during the decay of the outburst.

Figure \ref{fig:corona_sizes_flux} shows the evolution of the sizes of the two coronae with the flux ratio (left panel) and the QPO frequency (right panel), respectively. We found that both $L_1$ and $L_2$ decrease as the QPO frequency and the flux ratio increase. However, $L_2$ shows a higher scatter than $L_1$ in both cases.

\begin{table*}
\caption{Summary of the best-fitting parameters of the spectrum of the source and the rms and lag spectra of the type-C QPO in MAXI~J1348$-$630 in the 0.5--10 keV energy band. Errors represent the 1$\sigma$ level confidence interval of the parameter, while the upper limits represent the 95\% confidence upper bound of the corresponding parameter.}
\resizebox{\textwidth}{!}{
\begin{tabular}{ccccccccccc}
\hline 
 & Obs 1 & Obs 2 & Obs 3 & Obs 4 & Obs 5 & Obs 6 & Obs 7 & Obs 8 & Obs 9 & Obs 10 \\
\hline
\smallskip
\smallskip
$N_{\rm H}$ & 0.86$^*$ & 0.86$^*$ & 0.86$^*$ & 0.86$^*$ & 0.86$^*$ & 0.86$^*$ & 0.86$^*$ & 0.86$^*$ & 0.86$^*$ & 0.86$^*$ \\
\smallskip
\smallskip
$O$ & 0.77$^*$ & 0.77$^*$ & 0.77$^*$ & 0.77$^*$ & 0.77$^*$ & 0.77$^*$ & 0.77$^*$ & 0.77$^*$ & 0.77$^*$ & 0.77$^*$ \\ 
\smallskip
\smallskip
$F_{\rm e}$ & 0.34$^*$ & 0.34$^*$ & 0.34$^*$ & 0.34$^*$ & 0.34$^*$ & 0.34$^*$ & 0.34$^*$ & 0.34$^*$ & 0.34$^*$ & 0.34$^*$ \\ 
\smallskip
\smallskip
Flux$_{\rm disc}$ & (2.19 $\pm$ 0.05)$\cdot$10$^{-9}$ & (4.17 $\pm$ 0.05)$\cdot$10$^{-9}$ & (4.09 $\pm$ 0.07)$\cdot$10$^{-9}$ & (1.02 $\pm$ 0.02)$\cdot$10$^{-9}$ & (0.93 $\pm$ 0.02)$\cdot$10$^{-9}$ & (0.0511 $\pm$ 0.0006)$\cdot$10$^{-9}$ & (0.21 $\pm$ 0.01)$\cdot$10$^{-9}$ & (0.897 $\pm$ 0.004)$\cdot$10$^{-9}$ & (0.676 $\pm$ 0.008)$\cdot$10$^{-9}$ & - \\ 
\smallskip
\smallskip
$kT_{\rm in}$ & 0.465 $\pm$ 0.004 & 0.514 $\pm$ 0.003 & 0.328 $\pm$ 0.003 & 0.218 $\pm$ 0.002 & 0.201 $\pm$ 0.002 & 0.17 $\pm$ 0.01 & 0.239 $\pm$ 0.003 & 0.362 $\pm$ 0.001 & 0.349 $\pm$ 0.002 & 0.254 $\pm$ 0.006 \\ 
\smallskip
\smallskip
Flux$\rm _{Comp}$ & (26.2 $\pm$ 0.2)$\cdot$10$^{-9}$ & (27.6 $\pm$ 0.4)$\cdot$10$^{-9}$ & (7.60 $\pm$ 0.04)$\cdot$10$^{-9}$ & (5.29 $\pm$ 0.01)$\cdot$10$^{-9}$ & (4.71 $\pm$ 0.01)$\cdot$10$^{-9}$ & (4.79 $\pm$ 0.01)$\cdot$10$^{-9}$ & (6.44 $\pm$ 0.06)$\cdot$10$^{-9}$ & (8.68 $\pm$ 0.02)$\cdot$10$^{-9}$ & (7.99 $\pm$ 0.02)$\cdot$10$^{-9}$ & (8.71 $\pm$ 0.06)$\cdot$10$^{-9}$  \\ 
\smallskip
\smallskip
$\Gamma$ & 1.643 $\pm$ 0.004 & 1.571 $\pm$ 0.004 & 2.52 $\pm$ 0.03 & 2.347 $\pm$ 0.01 & 2.009 $\pm$ 0.007 & 1.675 $\pm$ 0.001 & 1.69 $\pm$ 0.003 & 1.691 $\pm$ 0.001 & 1.659 $\pm$ 0.002 & 1.691 $\pm$ 0.001  \\ 
\smallskip
\smallskip
$kT_{\rm e}$ & 50$^*$ & 50$^*$ & 50$^*$ & 50$^*$ & 50$^*$ & 50$^*$ & 50$^*$ & 50$^*$ & 50$^*$ & 50$^*$  \\ 
\smallskip
\smallskip
Index$_1$ & 1.49 $\pm$ 0.13 & 10. $^{+0.009}_{-0.43}$  & 4.05 $\pm$ 0.18 & 4.15 $\pm$ 0.08 & 2.91 $\pm$ 0.08 & 7.78 $\pm$ 0.02 & 7.42 $\pm$ 0.22 & 1.35 $\pm$ 0.06 & -10.$^{+7.89}_{-0.17}$  & 1.35 $\pm$ 0.01 \\ 
\smallskip
\smallskip
$a$ & 0.998 & 0.998 & 0.998 & 0.998 & 0.998 & 0.998 & 0.998 & 0.998 & 0.998 & 0.998  \\ 
\smallskip
\smallskip
$i$ & 49.417 & 49.417 & 49.417 & 49.417 & 49.417 & 49.417 & 49.417 & 49.417 & 49.417 & 49.417  \\ 
\smallskip
\smallskip
$\log x_{\rm i}$ & 2.48$\pm0.02$ & 42.1$\pm0.1$ & 0.330$\pm0.001$ & 3.29$\pm0.04$ & 22.9$\pm0.2$ & 68.8$\pm0.6$ & 91.7$\pm0.6$ & 41.1$\pm0.1$ & 2.47$\pm$0.08 & 2.0$\pm$0.2 \\ 
\smallskip
\smallskip
$A_{\rm Fe}$ & 3.28 $\pm$ 0.26 & 9.25$^{0.6}_{1.49}$ & 0.50 $\pm$ 0.02 & 0.50 $\pm$ 0.01 & 0.50 $\pm$ 0.02 & 3.670 $\pm$ 0.008 & 10.0$^{0.03}_{1.44}$  & 0.54 $\pm$ 0.01 & 0.50 $\pm$ 0.03 & 0.541 $\pm$ 0.002 \\ 
\smallskip
\smallskip
Norm$\rm_{ref}$ & 0.020 $\pm$ 0.003 & 0.042 $\pm$ 0.004 & 0.20 $\pm$ 0.02 & 0.099 $\pm$ 0.005 & 0.019 $\pm$ 0.001 & 0.005090 $\pm$ 0.000006 & 0.0052 $\pm$ 0.0002 & 0.0111 $\pm$ 0.0009 & 0.0098 $\pm$ 0.0002 & 0.011 $\pm$ 0.001  \\
\smallskip
\smallskip
$kT_{\rm s2}$ & 0.38 $\pm$ 0.03 & 0.26 $\pm$ 0.06 & 0.18 $\pm$ 0.02 & 0.17 $\pm$ 0.01 & 0.087 $\pm$ 0.003 & 0.092 $\pm$ 0.008 & 0.11 $\pm$ 0.01 & 0.13 $\pm$ 0.02 & 0.17 $\pm$ 0.03 & 0.228 $\pm$ 0.003 \\  
\smallskip
\smallskip
$L_{\rm 1}$ & 6313$^{+1404}_{-1149}$ & 764$^{+487}_{-252}$ & 45$^{+25}_{-15}$ & 1011$^{+279}_{-272}$ & 448$^{+131}_{-77}$ & 3548$^{+923}_{-567}$ & 2014$^{+873}_{-627}$ & 1798$^{+784}_{-599}$ & 742$^{+302}_{-196}$ & 5111$^{+2388}_{-1242}$  \\ 
\smallskip
\smallskip
$L_{\rm 2}$ & 39200$^{+357}_{-9303}$ & 28754$^{+239}_{-8406}$ & 662$^{+275}_{-220}$ & 44400$^{+3199}_{-8367}$ & 7275$^{+1257}_{-836}$ & 14500$^{+877}_{-578}$ & 62900$^{+16056}_{-10373}$ & 54800$^{+8277}_{-3085}$ & 21900$^{+1863}_{-7230}$ & 17433$^{+9288}_{-4047}$  \\ 
\smallskip
\smallskip
$\eta_{\rm 1}$ & 0.62$^{+0.2}_{-0.16}$ & 0.81 $\pm$ 0.16 & 0.67 $\pm$ 0.04 & 0.79 $\pm$ 0.08 & 0.50 $\pm$ 0.09 & 0.89 $\pm$ 0.02 & 0.74 $\pm$ 0.13 & 0.92 $^{+0.03}_{-0.11}$ & 0.96 $\pm$ 0.20 & 0.89 $\pm$ 0.03  \\ 
\smallskip
\smallskip
$\eta_{\rm 2}$ & $<$ 1.00 & $<$ 0.73 & 0.3 $\pm$ 0.04 & $<$ 0.03 & 0.27 $\pm$ 0.08 & 0.73 $\pm$ 0.02 & 0.99 $\pm$ 0.33 & 0.99 $\pm$ 0.13 & 0.55 $\pm$ 0.15 & 0.99 $\pm$ 0.01  \\ 
\smallskip
\smallskip
$\delta \dot H_{\rm ext1}$ & 1.59$^{+0.003}_{-0.19}$ & 0.53 $\pm$ 0.21 & 1.07 $\pm$ 0.27 & 1.06 $\pm$ 0.29 & 0.72 $\pm$ 0.14 & 0.78 $\pm$ 0.05 & 0.73 $\pm$ 0.15 & 0.17 $\pm$ 0.02 & 0.911 $\pm$ 0.27 & 1.49 $\pm$ 0.01  \\ 
\smallskip
\smallskip
$\delta \dot H_{\rm ext2}$ & 0.38$^{+0.1}_{-0.05}$ & 0.12 $\pm$ 0.05 & 0.84 $\pm$ 0.25 & 0.82$^{+0.09}_{-0.31}$ & 0.98$^{+0.005}_{-0.26}$ & 1.49 $\pm$ 0.07 & 2.46$^{+0.95}_{-0.74}$ & 0.29$^{+0.12}_{-0.27}$ & 0.403$^{+0.27}_{-0.04}$ & 0.72$\pm$0.11  \\ 
\smallskip
\smallskip
$\chi^2 / \nu$ & 249.21/196 & 227.39/197 & 233.85/192 & 221.84/196 & 242.76/190 & 205.58/192 & 220.44/190 & 888.78/852 & 230.09/197 & 211.42/196 \\
\hline
\end{tabular}}
\label{Tab:spec_params}
\end{table*}

\subsection{Broadband phase-lags and coherence vs. frequency}

\begin{figure}
    \centering
    \includegraphics[width=\columnwidth]{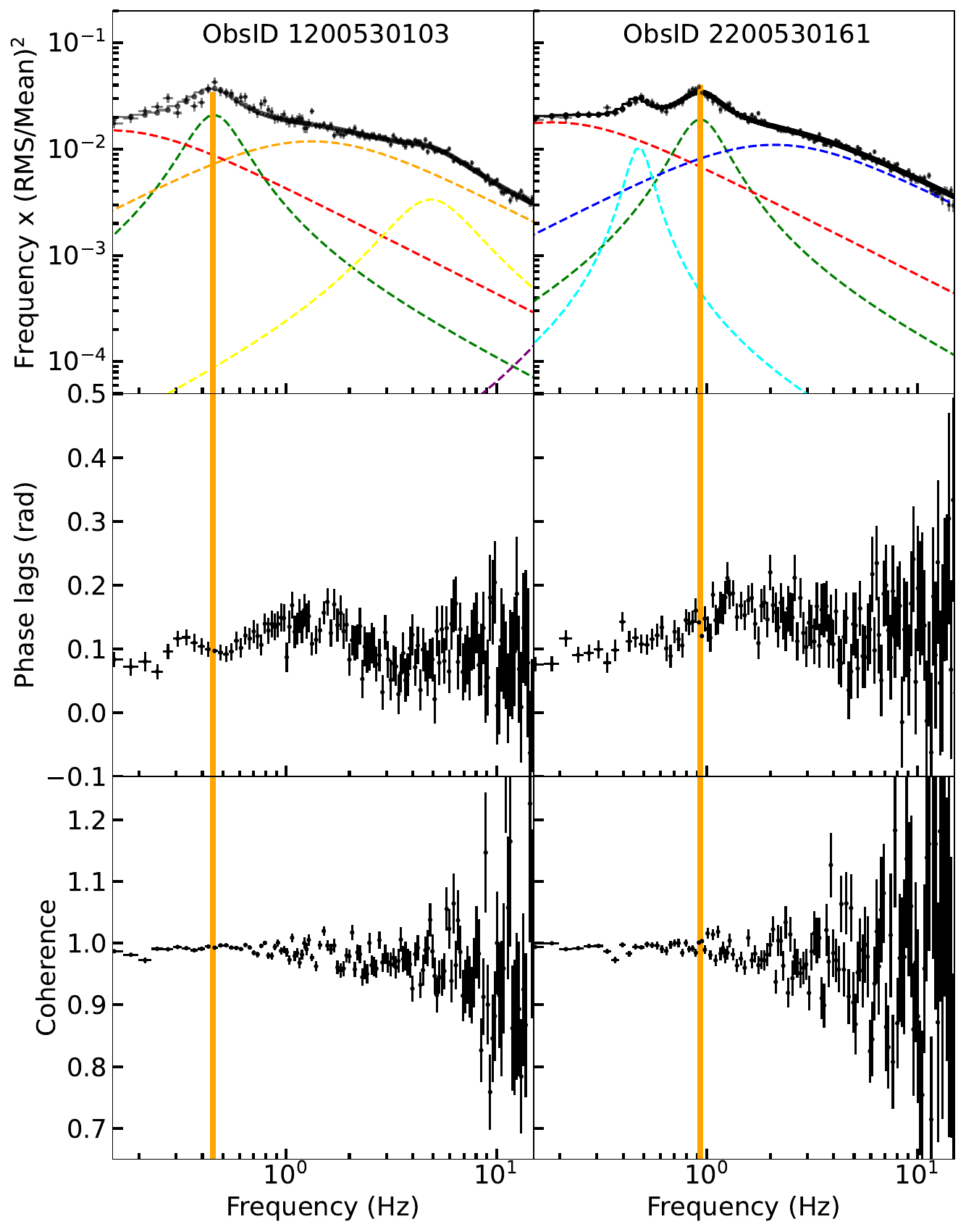}
    \caption{From top to bottom: PDS, phase lag and coherence vs frequency of the observations of MAXI~J1348$-$630 showing type-C QPOs during the rise of the main outburst (left column) and reflare (right column). The reference and subject bands are, respectively, 0.5$-$2.0 keV and 2.0$-$10.0 keV. Dashed lines represent the Lorentzians fitting the PDS. Green Lorentzian represents the type-C QPO, and the orange line marks the QPO frequency. Colours and symbols are as defined in the previous figure.}
    \label{fig:coherence_lhs}
\end{figure}

\begin{figure*}
    \centering
    \includegraphics[width=0.8\textwidth]{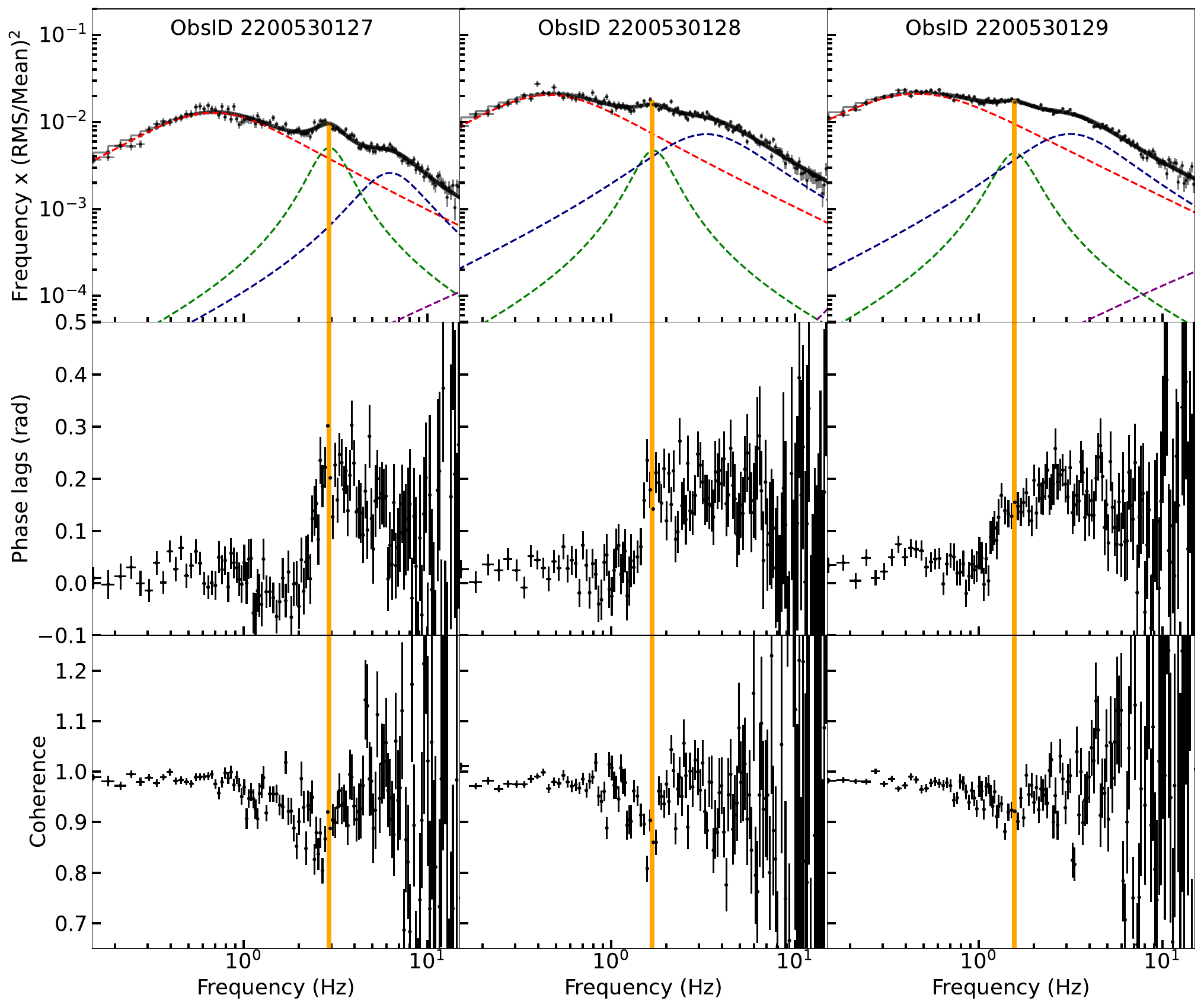}
    \caption{From top to bottom: PDS, phase lag vs frequency and coherence vs frequency of the observations of MAXI~J1348$-$630 showing type-C QPOs during the decay of the main outburst. The reference and subject bands, colours and symbols are as defined in the previous figure.}
    \label{fig:coherence_decay}
\end{figure*}

In \cite{Alabarta22}, we studied the properties of the type-C QPO observed in MAXI~J1348$-$630, and in the previous sections, we fitted the model \textsc{vkdualdk} to the fractional rms and phase lags spectra of the type-C QPO. Inspired by the recent works of \cite{Mendez24} and \cite{Konig24}, we decided to study the characteristics of the phase lags and the coherence function in two broad energy bands, 0.5$-$2.0 keV and 2.0$-$10.0 keV, vs. frequency of those observations showing a type-C QPO. Figures \ref{fig:coherence_lhs} and \ref{fig:coherence_decay} show the PDS (top panels), the phase lags (middle panels) and coherence (bottom panels) vs. frequency. During the rise of the main outburst and the reflare, both in the LHS, the broadband phase lags are more or less constant at $\sim$0.1 rad, with small excursions (less than 0.1 rad) at frequencies at which there is a feature in the PDS (Figure \ref{fig:coherence_lhs}, middle panels). At the same time, the coherence function is consistent with being $\sim$1 up to $\sim$10 Hz, where the errors increase significantly (Figure \ref{fig:coherence_lhs}, bottom panels). In particular, the coherence is $\sim$1 at the frequency corresponding to the type-C QPO (orange line in Figure \ref{fig:coherence_lhs}). On the contrary, during the decay of the main outburst, there is a sudden increase in the phase lags (Figure \ref{fig:coherence_decay}, middle panels) at the frequency of the QPO (also called ``the cliff''; see Bellavita et al. 2024, in preparation) while the coherence function (Figure \ref{fig:coherence_decay}, bottom panels) drops at that frequency (orange line in Figure \ref{fig:coherence_decay}). \footnote{\cite{Konig24} already showed this feature in ObsID 2200530129.} These sudden changes in both the lag and coherence are only observed in the three observations of the decay that show a type-C QPO in the PDS.

\section{Discussion}

We found that the rms and lag spectra of the type-C QPO detected in MAXI~J1348$-$630 with \textit{NICER} during its 2019 outburst and its reflare can be explained in terms of two independent but physically coupled Comptonization regions. We jointly fit the time-average spectrum of the source and the rms and phase-lag spectra of the QPO with the time-dependent Comptonization model \textsc{vkdualdk} \citep{Karpouzas20, Bellavita22} and deduced the size and the feedback fraction of the coronae in the three phases of the outburst (rise, decay and reflare). This is the first time that the geometry of the coronae has been studied during a reflare of an LMXB. During the LHS of the main outburst and during the reflare (where the source always remains in the LHS), the size of the small corona is of the order of $\sim$10$^{2}$$-$10$^{3}$ km with a feedback-fraction of $>$0.6. In comparison, the larger corona is of the order of $\sim$10$^{4}$ km with a high feedback fraction ($>$0.6). During the HIMS in the decaying part of the outburst, on the other hand, the large corona shows a lower feedback fraction ($\eta_2$ < 0.4). In addition, we found that the sudden increase of the phase lags with Fourier frequency, the so-called cliff, in the phase lag frequency spectrum, and the drop in the coherence function previously reported by \cite{Konig24} occur at the frequency of the type-C QPO in the three observations during the decay of the outburst.

\subsection{Geometry of the coronae}

During the rise of the main outburst (black circles in Fig. \ref{fig:corona_params_time}) and during the reflare (blue filled plusses in Fig. \ref{fig:corona_params_time}), the size of the small corona, $L_{1}$, ranges between $\sim$51$R_{g}$, (for a BH of 10$M_{\odot}$) and $\sim$425$R_{g}$. The large corona, on the other hand, shows sizes ranging from $\sim$980$R_{g}$ to $\sim$4250$R_{g}$.  
During the decay of the main outburst (red diamonds in Fig. \ref{fig:corona_params_time}), MAXI~J1348$-$630 was found to be in the HIMS \citep[e.g., ][]{Zhang20, Alabarta22}. In the three observations in this state where a type-C QPO was found, $L_1$ increases from $\sim$3$R_{g}$ to $\sim$70$R_{g}$, while $L_2$ shows values between 20$R_g$ and 1500$R_g$. 

In addition, the feedback fraction of both coronae provides a hint of their geometries. A high value of the feedback fraction suggests that the corona must be covering the accretion disc, such that a large fraction of the disc flux originates from the Comptonised photons that return from the corona to the disc. A possible geometry for this scenario is a corona extending horizontally parallel to the accretion disc. On the contrary, a low value of the feedback fraction implies that the disc flux produced by photons that impinge back from the corona is not very large. Two potential geometries can explain low feedback values. The first one is a small corona located around the compact object with a size $L$ smaller than the inner radius of the disc. The second one is a vertically extended corona, no matter its size, possibly moving away from the disc. During the rise of the main outburst and the reflare, both $\eta_1$ and $\eta_2$ range between $\sim$0.5 and $\sim$1.0. Considering that the sizes of the small and the large corona are, respectively, $L_1 \sim$10$^2$$-$10$^3$ $R_g$ and $L_2 \sim$10$^3$$-$10$^4$ $R_g$, we interpret that the two coronae are horizontally extended in the direction of the accretion disc, one being larger than the other one. The fact that the geometry is similar during both the rise of the outburst and the reflare reinforces the idea that both events are driven by the same physical mechanisms \citep[e.g., ][]{Patruno16, Cuneo20, Alabarta22, Saikia23}.
During the decay, the small corona presents a feedback fraction similar to that during the rise and the reflare, while the large corona shows a lower feedback fraction, ranging from $\sim$0 to $\sim$0.3. On the other hand, the sizes of both coronae are an order of magnitude smaller. This suggests that the geometry of the small corona is similar to that in the other two phases of the outburst, while the large corona is smaller and vertically extended. In all cases, the temperature of the seed photon of the large corona, $kTs_{2}$, is lower than the temperature of the inner part of the accretion disc, $kT_{in}$, suggesting that the large corona is illuminated by more external parts of the accretion disc, while the inner parts of the disc illuminate the small corona. In addition to this, the fact that $\eta_{int,1}$$\sim$0.03 during the rise and reflare and $\eta_{int,1}$$\sim$0.15 during the decay suggests that the disc is more truncated during the rise of the outburst and the reflare than during the outburst decay.

Taking into account the low values of $\eta_{int,2}$ from our fits, an alternative scenario for the large corona during the decay of the outburst would be the following: \citet[][Fig. 10]{Kylafis24} showed that only $\sim$$1-10$\% of the photons that escape at heights of 10$^{1}$-10$^{3}$ $R_g$ out of an outflowing corona of radius 100 $R_g$ at the base, and a Lorentz factor of 2.24, return to the disc. Assuming that $L_2$ in our fits represents the height at which the photons escape the outflowing corona, both 
the $L_2$ and $\eta_{int,2}$ values we found during the decay of the outburst ($L_{2}\sim$ 45$-$3000$R_g$ and $\eta_{int,2}\sim0.05$, respectively) are consistent with the results of \cite{Kylafis24}. Therefore, the large corona could be outflowing material \citep[e.g., ][]{Giannios04, Kylafis08, Reig18, Reig21, Kylafis24} and references therein.

Our results can be compared with those obtained with models of the family of \textsc{vkompth} for type-C QPOs observed in other BH~LMXBs \citep{Garcia22, Bellavita22, Zhang22, Rawat23, Rout23, Ma23}. \cite{Garcia22} and \cite{Bellavita22} fitted the fractional rms and phase lag spectra of the type-C QPO detected in GRS~1915$+$105 during the HIMS with a single-corona model. \cite{Garcia22} considered a blackbody as the seed photon source of the corona, while \cite{Bellavita22}, on the other hand, used a disk blackbody. Both studies obtained a corona with a size of $L \sim$10$^3-$10$^4$ km, $\eta <$ 0.2$-$0.3 and $kT_{s} <$ 0.2 keV, for QPO frequencies lower than 1.8 Hz (those consistent with the frequencies of the type-C QPO in MAXI~J1348$-$630). Due to the relatively large size and low feedback, they interpreted it as a vertically extended corona with a jet-like structure. 

\cite{Zhang22} and \cite{Rawat23} also modelled the fractional rms amplitude and phase lag spectra of the type-C QPO observed in MAXI~J1535$-$571 with \textsc{vkompth}. Using, respectively, \textit{HXMT} and \textit{NICER} data, these authors found a corona whose size decreases from $\sim$10$^4$ km to $\sim$10$^3$ km as the QPO frequency increases from $\nu_0$$\sim$1.8 Hz to $\nu_0$$\sim$2.7 Hz. The feedback fraction ranged between $\sim$0.6 and $\sim$0.8, meaning that a 10\%$-$25\% of the fraction of the corona flux returned to the disc. Because of this and the simultaneous behaviour of the source in radio frequencies, these authors suggested a geometry in which there is a small corona in the inner parts of the system and a larger corona vertically extended consistent with the jet. 

\cite{Rout23}, on the other hand, studied the geometry of the corona of GRO~J1655$-$40 fitting the type-B and type-C QPOs that were detected simultaneously during the 2005 outburst of the system in the so-called ultraluminous state. These authors fitted the spectrum of the source and the fractional rms and phase lag spectra of both QPOs simultaneously with a single comptonisation region. Their results indicate that the type-C QPO comes from a corona with a size of $L \sim$10$^3$ km and a feedback fraction of $\sim$0.6. These values suggested a horizontally extended corona covering partially the inner parts of the accretion disc. 

\cite{Ma23} studied the transition from type-C to type-B QPOs in MAXI~J1820$+$070 during, respectively, the HIMS and the SIMS. Both QPOs were fitted with \textsc{vkdualdk}, the model used in this study. For the type-C QPO, they found a small corona of $L_{1} \sim$74 km and an $\eta_{1} \sim$0.5, and a large corona of $L_{2} \sim$10$^4$ km and $\eta_{2} \sim$1.0. These parameters suggested a small corona located at the inner parts of the system and a large corona extending horizontally and covering the accretion disc, similar to that in GRO~J1655$-$40.

Except in GRO~J1655$-$40, the geometry of the corona in the above sources was studied during the HIMS. Comparing the results of MAXI~J1348$-$630 during the HIMS, we found similar results to those presented above. The small corona of MAXI~J1348$-$630 in the first observation of the decay of the main outburst is consistent with that of MAXI~J1820$+$070, both in size and feedback. After that, the corona of MAXI~J1348$-$630 increases in size. Regarding the large corona, both sources showed an $L_2 \sim$10$^3$ km, but while the feedback fraction in  MAXI~J1820$+$070 was $\sim$1.0, in MAXI~J1348$-$630 $\eta_2 < $0.4. While the interpretation of \cite{Ma23} was that the large corona extended horizontally in MAXI~J1820$+$070, we interpreted it as extending vertically in MAXI~J1348$-$630 due to the low feedback and the similar behaviour of the radio flux. The geometry of the large corona is more consistent with that of GRS~1915$+$105 presented in \cite{Garcia22} and \cite{Bellavita22}. The values of the corona in both GRO~J1655$-$40 and MAXI~J1535$-$571 are similar to those of the small corona of MAXI~J1348$-$630, although in MAXI~J1535$-$571
the corona is thought to be vertically extended. In MAXI~J1348$-$630 during the LHS, the sizes of both coronae are higher than those observed in other sources. However, this is expected. As we show in Fig. \ref{fig:corona_sizes_flux} and \cite{Rawat23} show in the top left panel of their Fig. 8, the size of the corona increases as the frequency of the QPO decreases. Then, the parameters we obtained are consistent with those described in the previous paragraph. We also note that independent results from X-ray polarization for Cyg~X$-$1 in the LHS and Swift~J1727.8$-$1613 in the HIMS show that the corona in the LHS is horizontally extended \citep{Krawczynski22, Ingram24}, as we found for MAXI~J1348$-$630. 

\subsection{Other geometries for coronae}

Previous works have studied the size and the evolution of the corona in BH~LMXBs using different geometries \citep[e.g., ][]{Ingram09, Marcel18_jed3, Marcel19_jed4, Kara19, Karpouzas20, DeMarco21, Karpouzas21, Garcia21, Marino21, Wang21_1820, Kawamura22, Mendez22, Wang22, Zhang23}. Assuming a lamppost geometry, \cite{Kara19} interpreted the evolution of the soft broadband lags of MAXI~J1820$+$070 and the constant width and centroid frequency of the $\rm Fe$ emission line as changes in the vertical size of the corona. In their interpretation the corona of MAXI~J1820$+$070 contracts in the LHS; in a subsequent paper \cite{Wang21_1820} proposed that the size of the corona increases again as the source evolves through the intermediate states. However, results from polarization studies suggest that the corona extends parallel to the accretion disc rather than vertically in the LHS \citep[e.g., ][]{Krawczynski22, Ingram24}, an interpretation that is also supported by the results of \cite{Ma23} and this paper. Alternatively, \cite{DeMarco21} interpreted the evolution of the soft broadband lags of MAXI~J1820$+$070 as a decrease of the inner disc radius instead of changes in the coronal geometry, which was assumed to remain constant. Assuming this is also true for MAXI~J1348$-$630, \cite{Zhang20} showed that the normalization of the disc blackbody increases during the rise of the reflare and decreases during the decay. Since the normalization of the disc blackbody is proportional to the inner radius of the disc squared, changes in the normalization suggest that the radius of the disc is also evolving, which would support the interpretation of \cite{DeMarco21}. However, our results fitting the \textsc{vkdualdk} strongly suggest that the size of the Comptonization region changes. Alternatively, in the JED-SAD model \citep{Ferreira97, Ferreira06_jed1, Petrucci08, Marcel18_jed2, Marcel18_jed3, Marcel19_jed4, Marcel20_jed5, Marcel22_jed6}, the corona is thought to be a jet-emitting disc (JED) linked to the standard accretion disc (SAD) through a transition radius ($r_j$). In \cite{Marcel20_jed5}, the frequency of the type-C QPO is linked with $r_j$ ($\nu_{QPO} \sim \nu_{K}$($r_j$)$/q$), where $\nu_{QPO}$ and $\nu_K$ are, respectively, the QPO and the Keplerian frequency, and $q=$133$\pm$4 is a scaling factor \citep{Marcel20_jed5}. If we take the frequency of the type-C QPO of MAXI~J1348$-$630, and assuming a 10$M_{\odot}$ mass for the compact object of the system, we get a transition radius ranging from 2$R_{g}$ to 7$R_{g}$. This transition radius is between 1 and 3 orders of magnitude lower than the characteristic sizes for both the coronae we inferred from the rms and phase-lag spectra of the type-C QPO of MAXI~J1348$-$630.

\subsection{Evolution of the coronae during the whole outburst and reflare of MAXI~J1348$-$630}

\begin{figure}
    \centering
    \includegraphics[width=\columnwidth]{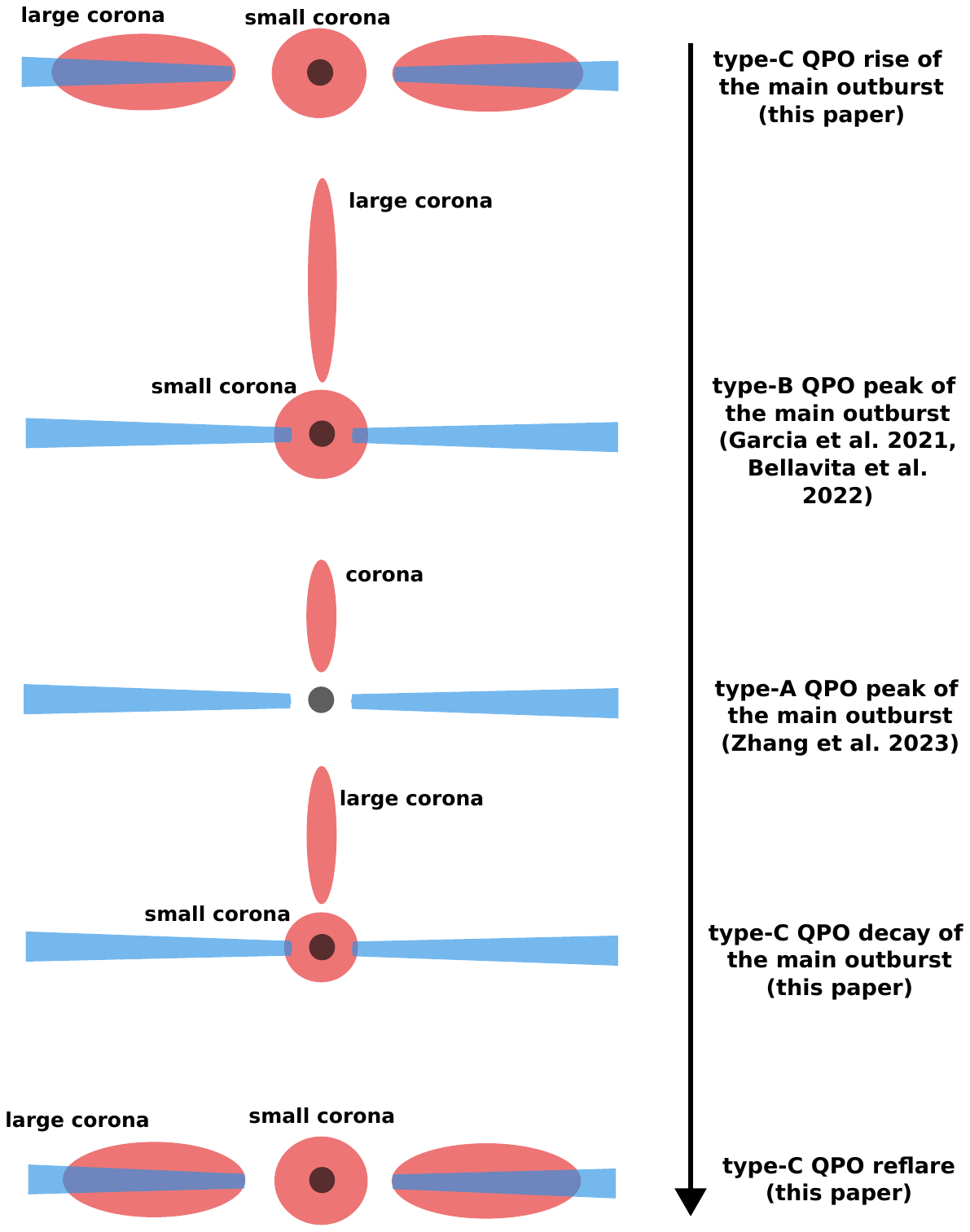}
    \caption{Schematic representation of the evolution of the coronae during the main outburst and the reflare of MAXI~J1348$-$630 through the study of the type-A, -B and -C QPOs.}
    \label{fig:schematic}
\end{figure}

Focusing on MAXI~J1348$-$630, the geometry of its coronae has also been studied by the analysis of the radiative properties of the type-B and type-A QPOs \citep[][]{Zhang21, Garcia21, Bellavita22, Zhang23_1348}. By carrying out a spectral-timing study of the fast appearance/disappearance of the type-B QPO, \cite{Zhang21} found that, when the type-B QPO is present in the PDS of the system, an increase in the Comptonization emission is detected. This increase can be well-fitted with an extra Comptonization component in the energy spectrum. Considering that this extra component only appears simultaneously with the type-B QPO, \cite{Zhang21} related it to the jet of the system. Independently, \cite{Garcia21} and \cite{Bellavita22} described the fractional rms and phase lag spectra of the type-B QPO by a model consisting of two physically connected coronae: a small and horizontally extended corona and a large one that is vertically extended. Both these authors used the version of the time-dependent Comptonization model for two coronae. \cite{Garcia21} used a blackbody component as a seed photon source and found a small corona with a size of $L_1 \sim$10$^3$ km and feedback fraction of $\eta_1 \sim$0.2 and a large corona with $L_2 \sim$10$^4$ km and $\eta_2 \sim$0.80. \cite{Bellavita22} used a disc-blackbody component as a seed photon source. In this case, the size of the large corona was estimated to be $L_2 \sim$13400 km with a relatively low $\eta_2 \sim$0.35. The small corona, on the other hand, presents a size of $L_1 \sim$160 km and a feedback fraction of $\eta_1 > 0.9$. They related the large corona with the emission of the base of the discrete jet that was being developed at the same time \citep{Carotenuto22}. \cite{Zhang23_1348}, on the other hand, jointly fitted the time-average energy spectrum of the source and the fractional rms and phase lag spectra of the type-A QPO with the model \textsc{vkompthdk}, that assumes a single corona with a disc blackbody as a seed photon source. They got a corona with a characteristic size of $\sim$2300 km and a $kT_s \sim$0.4 keV lower than $kT_{in}$. Due to the latter, they suggested a vertically extended geometry for the corona, similar to the one derived from the type-B QPO, but on a smaller scale. 

This paper completes the analysis of the \textit{NICER} data of the 2019 outburst of MAXI~J1348$-$630 with the family of \textsc{vkompth} models to fit the fractional rms and phase-lag spectra of the QPOs observed in this source. Therefore, we can compile the results derived from fitting the three different types of QPO and present the evolution of the properties of the coronae through the outbursts and the first reflare.

The first two observations in MAXI~J1348$-$630 (MJD 58511 and MJD 58512) with a QPO occur in the LHS during the rise of the main outburst and correspond to a type-C QPO of frequencies $\sim$0.4 Hz and $\sim$0.5 Hz. From the fits, we found two coronae horizontally extended: one small of $L_1 \sim$6300 km that shrinks to $L_1 \sim$760 km and a large corona that increases from $L_2 \sim$39200 km to $L_2 \sim$55000 km. In these two observations, the feedback fraction of the two regions is $\eta_1 \sim$0.7 and $\eta_2 <$1.0. Later, during seven days \citep[from MJD 58522 to MJD 58540, see ][ for the list of dates]{Belloni20} a type-B QPO was observed in the PDS of MAXI~J1348$-$630, and the source entered the SIMS. The small corona is still horizontally extended with a size of $L_1 \sim$150 km and $\eta_1 >$0.9, and the large corona is now vertically extended with a size of $\sim$12200 km and $\eta_2 >$0.25. Due to the behaviour of the system, the vertical corona could be associated as the base of the jet if the corona feeds the jet. A few days later \citep[from MJD 58543 to MJD 58539, see ][ for the list of dates]{Zhang23_1348}, when the type-A QPO was detected, the source was already in the HSS. Fitting the time-average spectrum of the source and the rms and phase-lag spectra of the QPO  with a single corona, a vertically extended corona of $\sim$2300 km and a feedback fraction of $\sim$0.5 was obtained. We can identify this as the large corona of the type-C and type-B QPOs that shrank in the HSS. When the decay of the outburst started, the source entered the HIMS again, and a type-C QPO was detected (from MJD 58603 to MJD 58606). The small corona, in this case, started with a size of $L_1 \sim$45 km, then increased to $L_1 \sim$1000 km and finally decreased to $L_1 \sim$400 km, with a feedback fraction ranging from $\sim$0.5 to $\sim$0.8. The large corona, on the other hand, started with a size of $L_2 \sim$600 km, then increased to $L_2 \sim$44000 km and finally decreased to $L_2 \sim$7000 km, with a feedback fraction always lower than $\sim$0.3. In MJD 58602 and MJD 58603, a sudden increase in the radio flux was detected, followed by a decrease in MJD 58607 \citep{Carotenuto21}. Because of this and the low $\eta_2$, we tentatively suggest that the large corona is associated with the base of the newly formed compact jet. Finally, during the reflare (see Table. \ref{Tab:observations}), the size of the small corona, $L_{1}$, ranges between $\sim$700 km and $\sim$3500 with a feedback fraction between $\sim$0.7 and $\sim$1.0. Extending the interpretation, the large corona became horizontally extended again, with sizes between $\sim$14500 km to $\sim$63000 km, with $\eta_2$ always higher than 0.5. A schematic representation of the evolution of the coronae is shown in Fig. \ref{fig:schematic}.

\subsection{Implications of the drop in coherence at the QPO frequency}

\cite{Konig24} found a sudden increase of the time lags and a sharp drop of the coherence as a function of Fourier frequency happening at frequencies between two broad Lorentzians in the PDS in some observations of Cyg~X$-$1, MAXI~J1820$+$070, MAXI~J1348$-$630 and AT~2019wey (see, e.g., their Fig. 17). In MAXI~J1820$+$070 and MAXI~J1348$-$630, both the increase of the lags and the drop in the coherence only appeared in the soft-to-hard transition happening during the decay of the outburst, and the frequency at which they are observed, increases as the source gets softer. Using a new analysis method, \cite{Mendez24} fitted the PDS and the CS with the same Lorentzian components and found that at the same frequency, a narrow Lorentzian appears in the CS but not in the PDS, being more significant in the Imaginary than in the Real part of the CS. Due to its large Imaginary part, \cite{Mendez24} called it ``Imaginary QPO''. This feature was interpreted as a potential beat between the two broad Lorentzians by \cite{Konig24}, whereas \cite{Mendez24} argued that Imaginary QPOs represent an independent phenomenon.

In the case of MAXI~J1348$-$630, the cliff in the phase lags and the drop in the coherence function occur at the same frequency as that of the type-C QPO. Contrary to the cases of Cyg~X$-$1 and MAXI~J1820$+$070, the features in the lags and coherence are consistent with a significant Lorentzian in the PDS. This is the first time that the cliff in the lags and the drop in coherence are linked to a component observed in the PDS and, in particular, to a type-C QPO.  Moreover, our result also suggests a potential relation between type-C and Imaginary QPOs, although a link between the latter and other types of QPOs cannot be ruled out (Bellavita et al. 2024, in preparation).  

This particular behaviour of the phase lags and the coherence function of MAXI~J1348$-$630 is only observed in the decay of an outburst during the soft-to-hard transition. This result is consistent with the Imaginary QPO of MAXI~J1820$+$070 \citep[][ Bellavita et al. 2024, in preparation]{Mendez24, Konig24}. The energy dependence of the phase lags of the type-C QPO during this phase of the outburst of MAXI~J1348$-$630 (Fig. \ref{fig:spec_params}) is very similar to that of the Imaginary QPO of MAXI~J1820$+$070 (Fig. 6 of Bellavita et al. 2024, in preparation). Both show large phase lag variations, of $\sim$0.8 rad and $\sim$1.5 rad, respectively, between soft and hard energies. In terms of the contribution of the Imaginary and Real parts of the CS, these larger lags reflect the fact that the cross vector has a large Imaginary part at the QPO frequency (see the definition of the lags in Section 2.3). The magnitude of the phase lag of the type-C QPO in the LHS of MAXI~J1348$-$630, on the other hand, is smaller (between $\sim$0.2 rad and $\sim$0.4 rad, see left and right panels of Fig. \ref{fig:spec_params}), meaning that, compared to the Real part, the Imaginary part of the CS is smaller than that in the HIMS.

On the basis of the above, two questions arise: why do the cliff in the lags and the drop in coherence only appear in the soft-to-hard transition and not in the LHS? And can we relate the behaviour of the lags and coherence to the evolution of the corona? We can conjecture about these questions under the light of the radiative properties of the type-C QPO of MAXI~J1348$-$630. Both in \cite{Alabarta22} and this paper, we showed that the fractional rms amplitude and the phase lags of the type-C QPO are consistent with being produced by oscillations in the properties of the corona. We also found evidence that the geometry of the corona of MAXI~J1348$-$630 changes through the outburst, similarly to other sources \citep[e.g., ][]{Kara19, Wang21, Garcia22, Mendez22, Rout23, Ma23}. When the source is in the LHS, the Comptonization region is described by two coronae: a small and a large one, both with relatively high feedback ($\eta_{1,2} >$ 0.5). Because of this, we interpreted them as being extended horizontally in the direction of the accretion disc. Since the feedback measures the fraction of the flux of the disc that originates from photons coming from the corona, a high feedback fraction implies a strong interaction between the corona and the accretion disc. This strong interaction can explain the high coherence ($\sim$1) that we observe at the frequency of the type-C QPO in Fig. \ref{fig:coherence_lhs}. On the contrary, the Comptonization region during the decay of the outburst, while the system is in the HIMS, is described by a small corona with similar properties as in the LHS and a large corona with a very low feedback fraction ($\eta_{2} <$ 0.3) that is interpreted to be vertically extended. This very low feedback implies that the large corona does not interact as much with the accretion disc as in the LHS. This weaker interaction between the large corona and the accretion disc could explain the drop in the coherence observed at the frequency of the type-C QPO during the decay of the outburst. Moreover, the different geometry of the large corona in the decay with respect to that in the LHS would explain why the drop in coherence is only observed during the state transition and not in the LHS.

\section{Summary and conclusions}

In this work, we show that the rms and lag spectra of the type-C QPO detected in MAXI~J1348$-$630 with \textit{NICER} during its 2019 outburst and reflare can be explained in terms of two independent but physically coupled coronae. Indeed, we can successfully fit jointly the time-averaged spectrum of the source and the rms and phase-lag spectra of the QPO with the time-dependent Comptonization model \textsc{vkdualdk} \citep{Karpouzas20, Bellavita22}. Moreover, we found a sudden increase of the phase lag frequency spectrum (the ``cliff'') and a drop of the coherence function at the same frequency of the type-C QPO during the decay of the outburst. This is the first time that this cliff in the lags and the corresponding sharp drop in the coherence function are directly connected to a type-C QPO in the PDS. We explain this cliff in the lag and drop in the coherence in terms of the geometry of the coronae of MAXI~J1348$-$630. Our main conclusions are the following:

\begin{enumerate}
    \item The fractional rms and phase lag spectra of the QPO can be fitted with two physically coupled coronae: one small and one large corona.

    \item During the rise of the main outburst and the reflare, both in the LHS, the characteristic size of the small corona is $\sim$10$^{2}$$-$10$^{3}$ km with a feedback-fraction of $>$0.6. The size of the large corona, on the other hand, is $\sim$10$^{4}$ km, and this corona also presents a high feedback fraction $>$0.6. Due to the high feedback fractions, the coronae are interpreted as being horizontally extended over the accretion disc.

    \item During the decay of the outburst, when the system is in the HIMS, the size of the small corona is $L_1 \sim$10$^1$$-$10$^3$ km and the feedback fraction is similar to that during the rise and the reflare. The size of the large corona is $\sim$10$^2$$-$10$^4$ km, with a lower feedback fraction ($\eta_2$ < 0.3). Due to the values of $\eta_2$ and the evolution of the radio flux during these observations, we interpreted the large corona as being vertically extended. 

    \item \cite{Konig24} found a sudden increase in the phase lag frequency spectrum (the cliff) and a drop in the coherence function in one observation during the decay of the outburst of MAXI~J1348$-$630. We found these features in the observations of the decay showing the type-C QPO but not in those with the QPO in the LHS in the rise and the reflare. Moreover, the cliff in the lags and the drop of the coherence occur at the same frequency as the QPO in the PDS, suggesting that this signal is responsible for these features. We hypothesise that the drop in the coherence function occurs in the HIMS but not in the LHS because of the differences in the geometry of large corona in both spectral states.

\end{enumerate}

\section*{Acknowledgements}

This material is based upon work supported by Tamkeen under the NYU Abu Dhabi Research Institute grant CASS. This work is based on observations made by the \textit{NICER} X-ray mission supported by NASA. This research has made use of data and software provided by the High Energy Astrophysics Science Archive Research Center (HEASARC), a service of the Astrophysics Science Division at NASA/GSFC and the High Energy Astrophysics Division of the Smithsonian Astrophysical Observatory. The authors acknowledge Dr. Sandeep Rout for the discussion about the geometry of the coronae. MM acknowledges the research programme Athena with project number 184.034.002, which is (partly) financed by the Dutch Research Council (NWO). FG is a CONICET researcher. FG acknowledges support by PIBAA 1275 and PIP 0113 (CONICET). YZ acknowledges support from the Dutch Research Council (NWO) Rubicon Fellowship, file no.\ 019.231EN.021.


\bibliography{bib_all}{}
\bibliographystyle{aasjournal}



\end{document}